\documentclass[12pt,twoside]{article}
\usepackage{amssymb}
\usepackage{amsmath}
\usepackage[makeroom]{cancel}
\usepackage{latexsym}
\usepackage{longtable}
\usepackage{epsfig}
\usepackage{graphicx,bbm,psfrag}
\graphicspath{{images/}}

\begin{document}
\begin{titlepage}
\begin{center}
\vspace*{2cm} {\Large {\bf  Fluctuating hydrodynamics for a chain of nonlinearly coupled rotators\bigskip\bigskip\\}}
{\large Herbert Spohn}\bigskip\bigskip\\
 Zentrum Mathematik and Physik Department, TU M\"unchen,\\
 Boltzmannstr. 3, D-85747 Garching, Germany\medskip\\
email:~{\tt spohn@ma.tum.de
}\end{center}
\vspace{5cm} \textbf{Abstract.}  We study chains of rotators from the perspective of nonlinear fluctuating hydrodynamics. As confirmed by previous MD simulations, at intermediate temperatures diffusive transport 
is predicted. At low temperatures we obtain the FPU scenario with suppressed heat peak.
\end{titlepage}
%%%%%%%%%%%%%%%%%%%%%%%%%%%%
%\section{Introduction}\label{sec1}
%\setcounter{equation}{0}
As recognized for some time, one-dimensional anharmonic chains may
have anomalous transport properties \cite{Livi97,LeLi03,Dh08,vB12}. For example, a small disturbance of the equilibrium state will spread
super-diffusively. But there is another large class of chains which exhibits regular diffusive transport.
Thus one central question is how to characterize 
these two universality classes. (There is also the class of integrable chains, as Toda and harmonic,
which have ballistic transport.) As one crucial feature, momentum conservation has been identified.
Momentum conservation holds for chains for which the interaction potential depends only on 
positional differences, the most prominent example being the Fermi-Pasta-Ulam (FPU) chains. On the other hand, if there is an on-site potential, then momentum is not conserved and the scattering from the underlying lattice potential forces regular transport.

A chain of nonlinearly coupled rotators seems to violate the above dichotomy. The interaction depends only on the  difference in angles. Thus angular momentum is conserved. Nevertheless, through molecular dynamics (MD) simulations diffusive behavior has been
convincingly demonstrated \cite{LeLi00}. The purpose of my note is to explain why, from the perspective of nonlinear fluctuating hydrodynamics, regular transport is indeed the natural option.

The hamiltonian of the rotator chain reads 
\begin{equation}\label{1}
H = \sum_{j=1}^N\big(\tfrac{1}{2}p_j^2 + V(\varphi_{j+1} - \varphi_j)\big)
\end{equation}
with periodic boundary conditions, $\varphi_{N+j} = \varphi_j$. The $\varphi_j$'s are angles and the $p_j$'s angular momenta. Hence the phase space is $([-\pi,\pi]\times\mathbb{R})^N$. The standard choice for $V$ is $V(\vartheta) =
-\cos \vartheta$, but in our context any $2 \pi$-periodic potential is admitted. The equations of motion are
\begin{equation}\label{2}
\frac{d}{dt} \varphi_j = p_j\,,\quad \frac{d}{dt}p_j = V'(\varphi_{j+1} - \varphi_j)- V'(\varphi_{j} - \varphi_{j-1})\,.
\end{equation}
Obviously angular momentum is locally conserved with the angular momentum current
\begin{equation}\label{3}
\mathcal{J}_1(j,t) = - V'(\varphi_{j}(t) - \varphi_{j-1}(t))\,.
\end{equation}
As local energy we define $e_j = \tfrac{1}{2}p_j^2 + V(\varphi_{j+1} - \varphi_j) $. Then $e_j$ is locally conserved, since
\begin{equation}\label{4}
\frac{d}{dt} e_j = p_{j+1} V'(\varphi_{j+1} - \varphi_{j})-p_jV'(\varphi_{j} - \varphi_{j-1})\,,
\end{equation}
from which one reads off the energy current
\begin{equation}\label{3a}
\mathcal{J}_2(j,t) = - p_j(t) V'(\varphi_{j}(t) - \varphi_{j-1}(t))\,.
\end{equation}

The hamiltonian for FPU type chains has the same structure as (\ref{1}),
\begin{equation}\label{5}
H_\mathrm{FPU} = \sum_{j=1}^N\big(\tfrac{1}{2}p_j^2 + V_\mathrm{FPU}(q_{j+1} - q_j)\big)\,,
\end{equation}
only now the positions $q_j \in \mathbb{R}$ and $V_\mathrm{FPU}$ should have at least a one-sided increase to infinity.
FPU type chains have three conservation laws, momentum, energy, and stretch $r_j = q_{j+1} - q_{j}$.
Also for angles one can define the difference $\tilde{r}_j = \varphi_{j+1} -\varphi_j \,\,\mathrm{mod} \,\,2 \pi$. Because of the modulo $2\pi$ such stretch is not conserved. A rotator chain has only two conserved fields.

According to nonlinear fluctuating hydrodynamics, the transport anomaly is related to the macroscopic Euler currents being nonlinear functions of the conserved fields \cite{Spohn14}. Thus for the rotator chain we have to compute the Euler currents in equilibrium.
Since there are two conserved fields the canonical equilibrium state reads
\begin{equation}\label{6}
\frac{1}{Z_N}\prod_{j= 1}^N \exp\big[ - \beta \big(\tfrac{1}{2}(p_j-u)^2 + V(\varphi_{j+1} - \varphi_j)\big)\big]  d\varphi_j dp_j\,,
\end{equation}
where $\beta > 0$ is the inverse temperature and $u \in \mathbb{R}$ the average angular momentum. Now
\begin{equation}\label{7}
\langle \mathcal{J}_1(j)\rangle_N = - \langle V'(\varphi_{j} - \varphi_{j-1})\rangle_N\,,\quad
 \langle \mathcal{J}_2(j)\rangle_N = - u\langle V'(\varphi_{j} - \varphi_{j-1})\rangle_N \,,
\end{equation}
average with respect to the canonical ensemble (\ref{6}). We claim that
\begin{equation}\label{8}
\lim_{N\to \infty}\langle V'(\varphi_{j} - \varphi_{j-1})\rangle_N = 0\,.
\end{equation}
For this purpose we expand in Fourier series as
\begin{equation}\label{9}
e^{-V(\vartheta)} = \sum_{m \in \mathbb{Z}}a(m) e^{-\mathrm{i} m\vartheta}\,,\quad f(\vartheta) = \sum_{m \in \mathbb{Z}} \hat{f}(m) e^{-\mathrm{i} m\vartheta}\,.
\end{equation}
Then, working out all Kronecker deltas from the integration over the $\varphi_j$'s, one arrives at
\begin{equation}\label{10}
\langle f(\varphi_{j+1} - \varphi_{j})\rangle_N  = \Big(  \sum_{m \in \mathbb{Z}}a(m)^N\Big)^{-1}
\sum_{m \in \mathbb{Z}}a(m)^N\Big(a(m)^{-1} \sum_{\ell \in \mathbb{Z}}\hat{f}(\ell -m)
 a(\ell)\Big)\,.
\end{equation}
Since $a(0) > |a(m)|$ for all $m \neq 0$,
\begin{equation}\label{11}
\lim_{N \to \infty} \langle f(\varphi_{j+1} - \varphi_{j})\rangle_N = a(0)^{-1} \sum_{\ell \in \mathbb{Z}}\hat{f}(\ell) a(\ell)   
= \frac{1}{Z_1} \int_{-\pi}^\pi d \vartheta f(\vartheta) e^{-\beta V(\vartheta)}\,.
\end{equation}
 For $f(\vartheta) = V'(\vartheta)$, the latter integral vanishes because of periodic boundary conditions in $\vartheta$.
 We conclude that both currents vanish on average.

On a large scale the conserved fields are expected to be governed by fluctuating hydrodynamics \cite{LL}.
Since the Euler currents vanish, the Langevin equations are linear with white noise currents and a dissipative second order drift
term.   There are two conserved fields and we form the equilibrium time correlations as
\begin{equation}\label{12}
S_{\alpha\alpha'}(j,t) = \langle g_\alpha(j,t)g_{\alpha'}(0,0) \rangle\,,\quad \vec{g}(j,t) = \big(p_j(t),e_j(t)\big)
\end{equation}
for the infinite lattice. $p_j$ is odd and $e_j$ is even under time reversal. Hence in the Green-Kubo formula
the cross term vanishes.  Thus, for large $j,t$,
\begin{equation}\label{13}
S_{\alpha\alpha'}(j,t) = \delta_{\alpha\alpha'}(4\pi D_\alpha t)^{-1/2}f_{\mathrm{G}}((4\pi D_\alpha t)^{-1/2}j )\,,
\end{equation}
where $f_{\mathrm{G}}$ is the unit Gaussian. $D_\alpha$ is the diffusion coefficient of mode $\alpha$.
Of course, it can be written as a time-integral over the corresponding total current-current correlation,
but its precise value has to be determined numerically. Energy diffusion is well confirmed in MD simulations \cite{LeLi00}. 
The total energy current correlation decays exponentially and the thermal conductivity depends on temperature as $e^{c_0 \beta}$ with some constant $c_0$ for sufficiently small temperatures. Momentum diffusion has been noted only recently \cite{Ha14}.

At low temperatures there is a more interesting scenario. If we assume that $V$ has a unique global minimum at some angle $\vartheta_0$, then the states of minimal energy are $\varphi_j  = j \vartheta_0 + \vartheta$,   $p_j = 0$. By redefining the angles one can always achieve $\vartheta_0 = 0.$ The angle
$\vartheta \in [-\pi,\pi]$ labels the broken rotational symmetry. Let us now fix some large $N$ and consider the initial state,
where $\varphi_1 = 0$ and $\varphi_{j+1} - \varphi_j$ are independent Gaussian random variables with variance
 $1/\beta V''(0)$, $V''(0) > 0$ by assumption. $\beta$ is taken so large that $\beta^{-1}N^{1/2}
 \ll 1$. This initial state is not stationary under the dynamics, but it has a very long life time. It is a rare event for the angle differences to change by $2\pi$. But then one can expand $H$ relative to the ground state configuration with the result
 \begin{equation}\label{14}
H = \sum_{j=1}^N\big(\tfrac{1}{2}p_j^2 + \tilde{V}(q_{j+1} - q_j)\big)\,.
\end{equation}
Here  $\tilde{V}$ is given by the Taylor expansion of $V(\vartheta)$ at $0$ up to the first even
power, larger than 2, which has a positive Taylor coefficient. On the low temperature scale, the deviations $
q_{j+1} - q_j$ take real values. Thus, in approximation, we have regained the three conservation laws of the FPU chain.

One can now apply the results from \cite{Spohn14}. The equilibrium time correlations have a three peak structure,
two symmetric sound peaks which move with the speed of sound and broaden according to the KPZ scaling function,
and a central heat peak which is standing still and broadens according to the $\tfrac{5}{3}$-Levy distribution.
At low temperatures the Landau-Placzek ratio for the heat peak is much smaller than the one for the sound peaks
\cite{St14}. Thus in a MD simulation one will see only the two sound peaks. At first, they  will move ballistically, but then slow down to eventually cross over to the Gaussian scaling (\ref{13}). The precise cross over still needs to be investigated.

In closing, we mention that the non-linear Sch\"{o}dinger equation on the one-dimensional lattice has  a  comparable structure. In this case the lattice field is $\psi_j \in \mathbb{C}$, where real and imaginary part are the canonically conjugate fields. The hamiltonian reads
\begin{equation}\label{15}
H = \sum_{j-1}^N\big(\tfrac{1}{2}|\psi_{j+1} - \psi_j|^2 + \lambda |\psi_j|^4 \big)
\end{equation}
with coupling $\lambda > 0$. This chain is non-integrable and the locally conserved fields are the number density
$n_j = |\psi_j|^2$ and the local energy $e_j = \tfrac{1}{2}|\psi_{j+1} - \psi_j|^2 + \lambda |\psi_j|^4$. The Euler currents vanish and both modes should have diffusive transport. This has been confirmed in \cite{Iu12,Iu13}, where also the two transport
coefficients are measured in their dependence on $\beta$ and the chemical potential $\mu$. However, at low temperatures
one observes sound peaks with KPZ scaling \cite{KuLa13,KuHuSp14,MeSp14b}. The mechanism explained before is at work. We fix the total number as $\sum_{j=1}^N |\psi_{j}|^2 = N$. In the ground state the 
global phase is broken and the minimizing field configuration equals $\psi^{\mathrm{G}}_j = \rho_0 e^{\mathrm{i}\vartheta}$ with $\rho_0 = 1$. Expanding
the hamiltonian (\ref{15}) at $\psi_j^{\mathrm{G}}$, momentum conservation is regained in approximation. The effective low temperature
hamiltonian is similar to (\ref{14}), but the leading correction to the Gaussian theory involves terms which couple the $p$'s and $q$'s. Stretch, momentum, and energy
are conserved, but the heat peak is hardly visible. The correlations have two symmetrically located  sound peaks which broaden according to KPZ scaling.

\end{document}